\documentclass[sigplan,10pt]{acmart}%
\acmJournal{PACMPL}
\acmVolume{1}
\acmNumber{PLDI} 
\acmArticle{1}
\acmYear{2020}
\acmMonth{1}
\acmDOI{} 

\setcopyright{none}

\usepackage[utf8]{inputenc} 
\usepackage{amsmath}
\usepackage{amsfonts}       
\usepackage{nicefrac}       
\usepackage{tabularx}
\usepackage{subcaption}

\usepackage[all,british]{foreign}

\usepackage[disable]{todonotes} 
\usepackage{pgfplots}
\pgfplotsset{%
    compat/show suggested version=false,
    compat=1.14,
}
\usepackage{pgfplotstable}

\usepackage{tcolorbox}
\tcbuselibrary{breakable}
\newtcolorbox{highlight}[1][]{%
    breakable,frame empty, sharp corners,
    boxrule=0pt,colback=yellow}

\usepackage{listings}
\renewcommand{\ttdefault}{pcr}
\makeatletter
\DeclareRobustCommand\ttfamily{\not@math@alphabet\ttfamily\mathtt\fontfamily\ttdefault\small\selectfont}
\makeatother

\newcommand{\codesize}{\fontsize{8}{7}\selectfont}

\newlength{\codeindent}
\lstdefinelanguage{SaC}[]{C}{%
    morekeywords={print,shape,dim,with,fold,modarray,genarray,+,-,=,*,
                  /,\%,sum,exp,min,drop,take},
    moredirectives={noinline},
}

\newcommand{\map}{\textit{map}}

\newcommand{\code}[1]{\texttt{#1}}
\providecommand{\shp}[1]{\code{shape}\,(#1)}
\newcommand{\tc}[3]{\{\,#1\;\text{->}\;#2\;|\linebreak\;#1\;\text{<}\;#3\,\}}
\newcommand{\sel}[2]{\code{#1}[\,\code{#2}\,]}
\newcommand{\meta}[1]{\texttt{\small <#1>}}

\begin{document}

\title{Array Languages Make Neural Networks Fast} 


\author{Artjoms {\v{S}}inkarovs}
\affiliation{
  \institution{Heriot-Watt University}   
  \streetaddress{Heriot-Watt University}
  \city{Edinburgh}
  \state{Scotland}
  \postcode{EH14 4AS}
  \country{UK}
}
\email{a.sinkarovs@hw.ac.uk}             

\author{Hans-Nikolai Vie{\ss}mann}
\affiliation{
  \institution{Heriot-Watt University}   
  \streetaddress{Heriot-Watt University}
  \city{Edinburgh}
  \state{Scotland}
  \postcode{EH14 4AS}
  \country{UK}
}
\email{hv15@hw.ac.uk}                    

\author{Sven-Bodo Scholz}
\affiliation{
  \institution{Heriot-Watt University}   
  \streetaddress{Heriot-Watt University}
  \city{Edinburgh}
  \state{Scotland}
  \postcode{EH14 4AS}
  \country{UK}
}
\email{s.scholz@hw.ac.uk}                

\newcommand{\cuda}{\textsc{CUDA}}
\newcommand{\tf}{\textsc{Ten\-sor\-Flow}}
\newcommand{\pt}{\textsc{Py\-Torch}}
\newcommand{\python}{\textsc{Python}}
\newcommand{\cc}{\textsc{C}}
\newcommand{\cxx}{\textsc{C++}}
\newcommand{\sac}{\textsc{SaC}}
\newcommand{\wloop}{\textsl{with}-loop}

\begin{abstract}
Modern machine learning frameworks are complex: they are typically organised
in multiple layers each of which is written in a different language and
they depend on a number of external libraries, but at their core they mainly
consist of tensor operations.  As array-oriented languages
provide perfect abstractions to implement tensor operations, we consider
a minimalistic machine learning framework that is shallowly embedded in an
array-oriented language and we study its productivity and performance.
We do this
by implementing a state of the art Convolutional Neural Network (CNN) and
compare it against implementations in TensorFlow and PyTorch --- two state of
the art industrial-strength frameworks.  It turns out that our implementation
is \emph{2 and 3 times faster}, even after fine-tuning the TensorFlow and
PyTorch to our hardware --- a 64-core GPU-accelerated machine.  The size of all
three CNN specifications is the same, about 150 lines of code.  Our mini
framework is 150 lines of highly reusable hardware-agnostic code that does not
depend on external libraries.  The compiler for a host array language
automatically generates parallel code for a chosen architecture.  The key to
such a balance between performance and portability lies in the design of
the array language; in particular, the ability to express rank-polymorphic
operations concisely, yet being able to do optimisations across them.  This design
builds on very few assumptions, and it is readily transferable to other
contexts offering a clean approach to high-performance machine learning.

\end{abstract}



\maketitle

\section{Introduction}

With the increasing success of machine learning in various domains, scientists
attempt to solve more and more complex problems using neural networks and deep
learning.  Increased complexity in the context of deep learning typically
means more layers of neurons and larger training sets all of which results
in the necessity to process larger amounts of data.
As a result modern networks require advanced and powerful hardware ---
modern machine learning applications are envisioned to
run on massively parallel high-throughput systems that may be equipped
with GPUs, TPUs or even custom-built hardware.

Programming such complex systems is very challenging, specifically in an
architecture-agnostic way.  Therefore, there is a big demand for a system
that abstracts away architectural details letting the users to
focus on the machine learning algorithms.
\tf{} or \pt{} solve exactly that problem --- they provide a convenient level
of abstraction, offering a number of building blocks that machine learning
scientists can use to specify their problems.  Productivity of such a solution
is quite high as these frameworks are embedded into
high-level languages such as \python{} or \cxx{}.

However, turning a framework-based specification into an efficient code remains
challenging.  There is a huge semantic gap between the specification
and the hardware, yet frameworks such a \tf{} and \pt{} introduce many
levels of abstractions that one needs to deal with.  Typically there is
a \python{} front-end, a core library in \cxx{} that depends on numerous external
libraries for linear algebra, tensor operations, libraries for GPUs and
other specialised hardware.  Such a complexity makes it challenging to
deliver excellent performance: optimisations across multiple layers of
abstraction as well as across multiple external libraries inherently come with
overheads.

The key question we are investigating is: can we identify a single layer
of abstraction where on the one hand we can express the core building blocks
and generate efficient parallel code, and on the other hand that is high-level
enough to be used as a front-end.

Based on the observation that neural networks can be concisely expressed as
computations on high-ranked tensors, we look into using a shape-polymorphic
array language \ala{} APL~\cite{Iverson:1962:APL} as the central layer of abstraction.
While APL itself seems suitable in terms of expressiveness~\cite{Sinkarovs:2019:CNN:3315454.3329960}
the interpreter-based implementation of operators, non surprisingly, does not readily provide
parallel performance anywhere near that of \tf{} or \pt{}.

However, over the last 50 years we have seen quite some research into compilation
of array languages into efficient parallel code~\cite{Cann:1990:SVF:110382.110593,RBernecky:apexinternals,sac,henriksen2017futhark,DBLP:conf/cgo/SteuwerRD17}.
These languages leverage whole program optimisations and they offer
decent levels of parallel performance.  They also offer high
program portability, as inputs are typically hardware-agnostic and
all the decisions on optimisations and code generation are taken by
the compiler.  A user can influence these decisions by passing options,
but no code modifications are required.


For the purposes of this paper we use \sac{}~\cite{sac}, a functional
array language, as our implementation vehicle.  We focus on a simple yet frequently
benchmarked CNN for recognising handwritten characters.  First we
implement building blocks that are required to define a chosen CNN in native
\sac{} and then we use these building blocks to define the network.
We compare the resulting code size and performance against \tf{} and \pt{}.
We observe that the overall problem can be expressed
concisely (300 lines of native\footnote{The code does not depend on any
specialised numerical libraries like MKL, only system libraries like libc or
pthreads.} SaC code)
and on a GPU-accelerated 64-core machine, our solution
performs \emph{two and three times faster} than the \tf{}-
and \pt{}-based implementations.
The key aspect of such good performance is first-class support for
multi-dimensional arrays in a functional setting followed by a number of
well-known code-generation techniques used by the chosen compiler.

This example suggests that at least for this particular domain, the trade-off
between conciseness, performance and development time is quite satisfying.

The individual contributions of the paper are:
\begin{itemize}
    \item we make a case for using array languages to host a machine-learning framework,
    \item we provide a concise implementation of the CNN for hand-written image
        recognition in \sac{} without using any domain-specific libraries, and
    \item we present a performance evaluation of the CNN in \sac{} against multiple variants
        of the \pt{}- and \tf{}-based versions of the algorithm on a
        high-performance cluster node.
\end{itemize}

The rest of the paper is organised as follows.  In Section~\ref{sec:background}
we briefly introduce machine learning algorithms and state of the art
frameworks.  In Sections~\ref{sec:sac-arrays}~and~\ref{sec:cnn} we introduce
the notion of functional arrays and describe our implementation of the CNN\@.
Section~\ref{sec:performance} presents performance and productivity evaluation.
Section~\ref{sec:related} reviews related work, and we conclude in
Section~\ref{sec:conclusions}.

\section{\label{sec:background}Background}

In the last decade machine learning attracted a lot of attention as it offered
solutions to several practical problems that mainly have to do with automatic
recognition of complex patterns: objects in images or videos, automatic
text translation, recommendation systems, speech recognition, \etc{}
Due to space limit we only focus on the computational aspects of machine
learning algorithms and CNNs in particular.
For an in-depth review refer to~\cite{schmidhuber2015learning,INDOLIA2018679}.

All machine learning algorithms
are based around the idea that we want to learn
(through \emph{guessing}) the function $f$ that maps the input variables $X$ to the output variables
$Y$, \ie{} $Y = f\ X$,  in the best possible way, according to some cost function.
After $f$ is found for the existing sample, we would like to make new predictions for
new inputs.

\paragraph{Linear Regression}
The simplest example of a machine learning algorithm is linear
regression~\cite{gauss1809theoria,legendre1805nouvelles}.
It is probably one of the most well-understood algorithms in the area, yet it
demonstrates fundamental principles that will be also used in CNNs. 
Given a set of $n$ statistical units $\{y_i\;,\; x_{i1}, \dots, x_{im}\}$, for
$i \in \{1, \dots, n\}$, we assume that the relationship between $y$s and $x$s
is linear, so that each ${y}_i$ can be computed as:
\(
    {y}_i = \beta_0 + \beta_1 x_{i1} + \cdots + \beta_m x_{im} + \epsilon_i
\).
This can be written in matrix form as:
\begin{gather*}
    {y} = X\beta + \epsilon\qquad\text{where}\\
    {y} = \begin{pmatrix}
        {y}_1\\
        \vdots\\
        {y}_n
        \end{pmatrix}
    \quad
    X = \begin{pmatrix} 
        1 & x_{11} & \cdots& x_{1m} \\
        \vdots & & \ddots \\
        1 & x_{n1}& \cdots& x_{nm}
        \end{pmatrix}
    \quad
    \beta = \begin{pmatrix}
            \beta_0\\
            \vdots\\
            \beta_m
        \end{pmatrix}
    \quad
    \epsilon = \begin{pmatrix}
                \epsilon_1\\
                \vdots\\
                \epsilon_n
            \end{pmatrix}
\end{gather*}
There exists a large number of methods to estimate or infer parameters 
$\beta$ and $\epsilon$ such that our model function ``best'' fits the
data.  For example, one commonly used method is \emph{linear least
squares}~\cite{legendre1805nouvelles}.  We assume that $\epsilon = 0$ and the cost function that we
want to minimise is:
\(
    \sum_{i = 1}^n{(y_i - \hat{y}_i)}^2
\)
where
\(
    \hat{y}_i = X\beta
\).
The attractiveness of this method lies in existence of the closed solution for
the parameter vector $\beta$ given by the formula:
\(
    \beta = {(X^{\top} X)}^{-1}X^\top y
\).

Note two important aspects.  First,
instead of searching through all the functions from $X$ to $Y$, we restrict the
general shape of that function and introduce a set of parameters ($\beta$-s in
our case).  The search of a function reduces to the search of the parameters.
Secondly, computationally,
most of the involved operations can be reduced to linear algebra operations.
This means that we will need a representation for vectors, matrices, tensors
and common operations on them when implementing machine learning algorithms.

\paragraph{Neural Networks}
\todo[inline]{need to make this flow with above}
Continuing on from linear regression, we can consider that the function $f:x\to Y$
that we want to learn as a composition of functions $g_i$ that can be further decomposed
into smaller functions. Overall such a composition forms a graph (or 
\emph{network}) connecting inputs $X$ with outputs $Y$.
 
A typical function composition takes the form:
\(
    f\ x = A\ \left(\sum_i w_i\ (g_i\ x)\right)
\)
where $A$ is an activation function (usually it is chosen to be continuous and
differentiable, \eg{} sigmoid, hyperbolic tangent, \etc{}) and $w_i$ are
so called weights.  These weights are parameters of our approximation that
we want to find, similarly to $\beta$ in linear regression, so that our
cost function is minimised.

Usually, neural networks are designed in a way that offers slicing of the
elementary functions $g_i$ into layers, so that all the elements in the given
layer can be computed independently.  As a layer is an activation function
of the weighted sum of other layers, most of the transitions in the network
can be expressed as matrix or tensor operations.

Very often due to the size and complexity of the network, the closed
solution that finds optimal weights either does not exist or is very difficult to
find.  Therefore, weight prediction is usually performed in an iterative manner.
In this case, the concept of the backpropagation --- a method to calculate
the gradient of the objective function with respect to the weights, becomes of a
significant importance.  On the one hand it provides a working solution that
is straight-forward to compute:
\(
    w := w - \eta \nabla F(w) 
\)
where $w$ are all the weights in the given network.  In the cases when our
objective function can be written as: $F = \sum_i F_i$, the gradient descent
can be rewritten as: $w - \eta\nabla\sum_i F_i = w - \eta\sum_i\nabla F_i$.
Furthermore, the stochastic gradient descent~\cite{SGD} approximates the true
gradient as follows:
\(
    w := w - \eta\nabla F_i(w)
\)
which is typically more efficient.  Intuitively, if we process a batch of
items, we can update weights after processing one individual item.
Finally, with carefully chosen activation functions $A$,
the computation of the backpropagation can be expressed as
a composition of linear-algebraic operations.

\paragraph{Chosen Problem}
CNNs~\cite{schmidhuber2015learning,INDOLIA2018679}, are neural networks where at least
one layer is computed as a convolution of the values from the previous layers. In this paper
we will implement a CNN and use it to recognise
hand-written digits. We base our implementation on Zhang's network design~\cite{zhang2016derivation}.
For training and recognition we rely
on the widely used MNIST data set\footnote{see \url{http://yann.lecun.com/exdb/mnist/}.} as input.

\paragraph{State of the Art Machine Learning Frameworks}
The overall design of state of the art machine learning
frameworks such as \tf{}~\cite{tf-intro}, Caffe~\cite{caffe-intro},
CNTK~\cite{cntk-intro}, Torch~\cite{torch-intro}, or \pt{}~\cite{pytorch-intro}
are very similar.  There is a core part written in \cc{}/\cxx{} with the use
of external libraries, and there is an interface part --- usually a Python library.
The core part contains highly-optimised kernels doing
tensor operations, linear algebra operations, and convolutions, that are
pre-optimised for the range of supported architectures.  All these frameworks
support computations on the GPUs, multi-threaded and distributed executions.
\tf{} also supports custom hardware known as Tensor Processing Units (TPU).

The main difference between the frameworks lies in the number of building blocks that they
provide which in turn influences the productivity of data scientists.  For
instance, Caffee and CNTK make it possible to specify networks via a configuration
file allowing users to avoid programming entirely.  Differences in the underlying
libraries (BLAS, tensor libraries, GPU libraries) and optimisation techniques
(XLA compiler, just-in-time compilation, kernel fusion) lead to runtime differences
on the chosen hardware.

All frameworks have in common that they
construct an internal representation of the dataflow graph of the network.  
This representation makes it possible to support automatic differentiation which
automates the computation of gradient descents.
Furthermore, such dataflow graphs are being analysed in  order to exploit natural concurrency of the
network, optimise the scheduling of multiple network nodes across the available devices or threads, \etc{}
In \tf{} and CNTK the graph is statically fixed, whereas in \pt{} the
graph can change at runtime.

\paragraph{\label{sec:sac-arrays}The Essence of Array Programming}

The underlying linear algebra of CNNs suggests that any implementation
is amenable to a formulation based on multi-dimensional arrays.
\todo[inline]{mention NumPy and Eigen here?}
Any declarative array language as powerful as APL, the $\Psi$-calculus, or
\sac{}\@ can be used to express tensor operations.

Conceptually, all that is needed is an abstraction for $n$-dimensional arrays, with three basic primitives:
selection, shape-enquiry and some form of
$n$-dimensional map functionality.
%
%
In \sac{}~\cite{sac,sac-mt}, arrays can be constructed by using
square brackets:
\begin{gather*}
    a = [1,2,3]\qquad b = [[1,2],[3,4],[5,6]]\\
    c = [[[1,2],[3,4]], [[5,6],[7,8]]]
\end{gather*}
It is assumed here, that all arrays are rectangular, \ie{}
all nestings are homogeneous, and expressions like $[[1,2],[3]]$
are considered ill-formed.
Each array has a \textsl{shape} which is a vector (1-dimensional array)
denoting the number of elements per axis.
For the above examples, we have:
\begin{align*}
    \shp{a} = [3] & \quad\shp{b} = [3, 2] \\
    \shp{c} &= [2,2,2]
\end{align*}
All expressions are considered arrays --- empty arrays as
well as scalar values also have shapes:
\begin{gather*}
    \shp{[]} = [0]\qquad \shp{[[]]} = [1, 0]\\
    \shp{42} = []
\end{gather*}
The shape
of an empty vector is $[0]$;  the shape of the 2D array containing one row that contains
no elements is $[1,0]$, and the shape of a scalar value
is the empty vector.
Selections have C-like syntax \sel{\meta{array}}{\meta{iv}} (where \meta{iv} is shorthand for \meta{index-vector})
and the following two constraints:
\begin{enumerate}
    \item $\shp{\meta{iv}} \leq{} \shp{\shp{\meta{array}}}$\\
        the length of the index vector can at most be as long as the array has axes, and
    \item $\meta{iv} \mathop{\dot{<}} \shp{\meta{array}}$\\
        the values of the index vector must be in range, \ie{} element-wise less
        ($\dot{<}$) than the corresponding shape elements.
\end{enumerate}
In case \meta{iv} has maximal length, the corresponding scalar element
in \meta{array} is selected.
Otherwise, the selection pertains to the first axes of \meta{array} only and
returns a sub-array whose shape corresponds to those components of the
shape of \meta{array} for which no indices were provided. In case \meta{iv} is empty, the entire array is selected.

Finally, \sac{} provides a data-parallel array constructor for $n$-dimensional
arrays named \wloop{}. For the context of this paper, we use its
shorthand notation that we call \emph{array comprehension}.  An $n$-dimensional array can be specified by an expression of the form:
\[
    \tc{\meta{idx-var}}{\meta{elem-expr}}{\meta{shp-expr}}
\]
where the shape of the result is determined by the value of $\meta{shp-expr}$,
and each element is computed by evaluating the expression $\meta{elem-expr}$.
\sac{} allows $\meta{elem-expr}$ to evaluate to non-scalar arrays,
provided that all these expressions are of identical shape.  The shape of the
overall result is the concatenation of $\meta{shp-expr}$ and $\shp{\meta{elem-expr}}$.
For example, we have:
\begin{align*}
    \tc{\code{iv}}{1}{[3]} &= [1,1,1] \\
    \tc{\code{iv}}{[1,2]}{[2]} &= [[1,2],[1,2]]
\end{align*}
The index variable can be referred-to in the element expression, \eg{}
an expression of the form \tc{\code{iv}}{\code{a}[\code{iv}]+1}{\shp{a}} computes an array
that has the same shape as a given array \code{a} but whose elements have been
increment by one.  This notation is an extended version of the set-expressions in~\cite{IFLmadrid};
it has been implemented in the latest version
of the \sac{} compiler and will be available in the next release.

\paragraph{More on \sac{}}
We capture the set of assumptions in \sac{} that enable a compiler to
generate efficient code.  Firstly, \sac{} is a first-order functional language.
This means that all the functions are pure, and all data is immutable.
Conceptually, every assignment copies its right hand side and every function
call copies its arguments.
Such an assumption makes memory management completely transparent --- there
is no way to force a memory allocation, and there is no way to pass a pointer.
The concept of pointers and references does not exist as it would break
the assumption about purity.  This makes all the optimisations much
simpler as there is no need to solve the aliasing or ownership problems.
At runtime we avoid copying data that can be shared with the help of
reference counting.

Secondly, the \sac{} compiler has multiple backends for generating
code for sequential, multi-threaded and CUDA architectures from a single
specification.  No user annotations are needed to indicate parallel regions,
as the \wloop{} per semantics exposes parallelism. Given that every
iteration can be run concurrently, the compiler chooses which array comprehension
will be run in parallel and generates either a multi-threaded version of the
code or a CUDA kernel.

Finally, \sac{} uses C-like syntax for functions and comes with a rich standard
library of pre-defined array operators.

\section{\label{sec:cnn}CNN}

In this section we describe our implementation of a CNN using~\cite{zhang2016derivation}
as a blueprint. It constitutes a typical CNN for recognising handwritten images of digits.
\begin{figure*}[th]
\includegraphics[width=.95\textwidth]{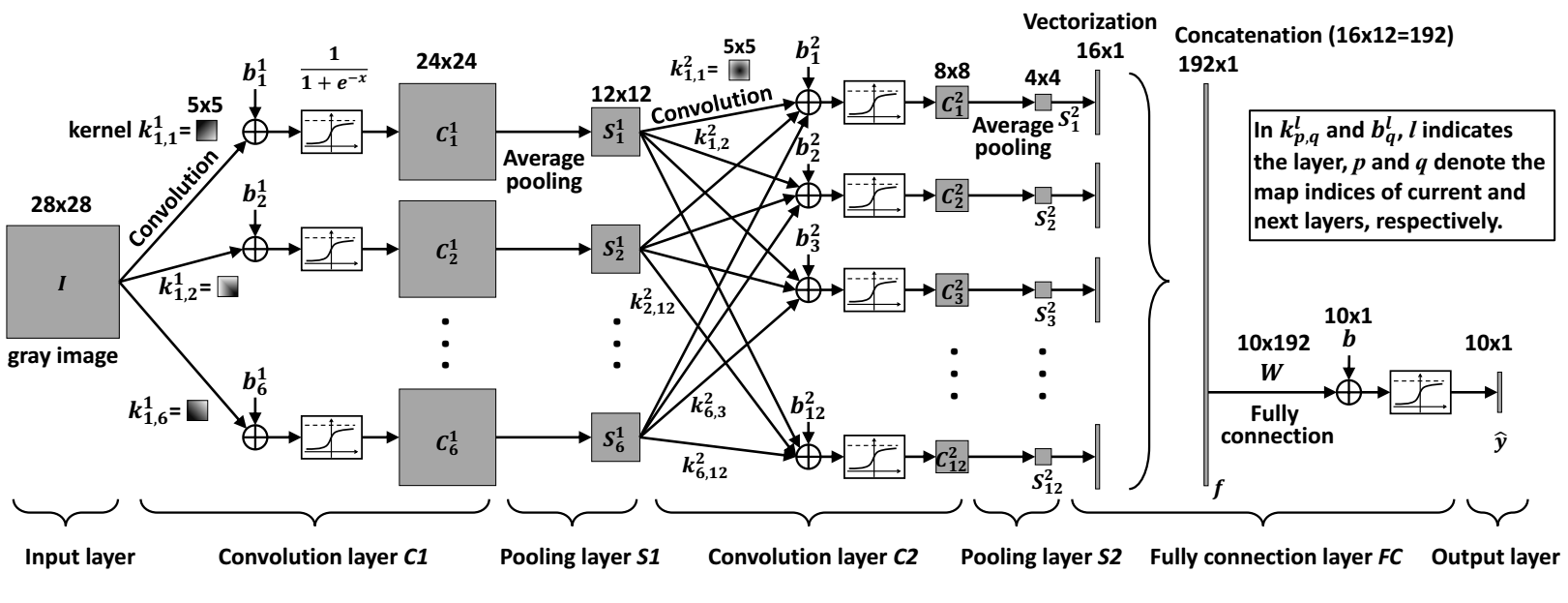}
\caption{\label{cnn}CNN for digit recognition. \textit{The picture is taken
        from~\cite{zhang2016derivation}}}
\end{figure*}
\hyperref[cnn]{Figure~\ref*{cnn}} shows the construction of the network which starts from
a $28\times28$ pixel image of a digit and produces a 10-element $\hat{y}$
through a sequence of convolution and pooling layers.  The vector $\hat{y}$
contains the probabilities of the input actually depicting the digits 0--9.

\paragraph{Convolution}
In the first layer $C_1$, we compute six convolutions of the input image $I$
with $5\times5$ matrices of weights $k^1_{1,i}$ producing six $24\times24$
arrays.  One such convolution can be implemented as:
\begin{lstlisting}[language=sac]
float[*] conv(float[*] I, float[*] k) {
  return { iv -> sum({ ov -> I[iv+ov] * k[ov]
                     | ov < shape(k) })
         | iv < shape(I) - shape(k) + 1 };
}
\end{lstlisting}
The type \code{float}[*] denotes an array of floating
point numbers of arbitrary shape.  For our image $I$ of shape $[28,28]$
and any of the weights $k^1_{1,i}$ of shape $[5,5]$, the result
is of shape $[28,28] - [5,5] + 1 = [24,24]$.
Each element at the index position \code{iv} is computed as a sum
of $5\times5$ elements in \code{I} multiplied with the corresponding weights in \code{k}.

Using \code{conv} we can define a function \code{mconv} to compute the six convolutions
and to add the individual biases $b^1_i$
to each convolution (denoted as $\oplus$ in \autoref{cnn}):
\begin{lstlisting}[language=sac]
float[*] mconv(float[*] I, float[*] k, float[*] b) {
  return { i -> conv(I, k[i]) + b[i]
         | i < shape(b)};
}
\end{lstlisting}
This function is rank-polymorphic, and in the context of
the $C_1$ layer we chose to store all $k$s in a 3D array of shape
$[6,5,5]$.  One bias per convolution leads to the shape of \code{b} being $[6]$.
For every index in \code{b}, \code{mconv} computes the convolution
of $I$ with \sel{k}{i} that is adjusted by adding the \sel{b}{i} bias.
The expression \sel{k}{i} selects a $[5,5]$ sub-array at the corresponding
index, and `\texttt{+} \sel{b}{i}' adds the scalar to every element of the $[24,24]$
array, resulting in the overall shape $[6,24,24]$.

The last step in $C_1$ is the application of the sigmoid activation function
to all values. We define it with the overloaded versions of mathematical
functions provided in the standard library of \sac{}:
\begin{lstlisting}[language=sac]
float[*] sigmoid (float[*] in) {
  return 1f / (1f + exp(-in));
}
\end{lstlisting}
This is a rank-polymorphic shape-preserving function, so its application
to the result of \code{mconv} of shape $[6,24,24]$ yields the desired result of
the same shape.

Consider now the convolution layer $C_2$.  If we choose to represent $s$ as a 3D
array of shape $[6,12,12]$, our rank-polymorphic specification of \code{mconv}
becomes immediately applicable here.  Intuitively, if $s$ is a single object,
then all the left hand sides of the arrows from $S_1$ to $C_2$ in \autoref{cnn}
will merge into a single point, similarly to the first convolution.  Our new
input to \code{conv} is of shape $[6,12,12]$, so each $s^i$ should be of shape
$[6,5,5]$, producing a result of shape $[1,8,8]$.  As we have 12 $s^i$ and 12
biases, the application of the \code{mconv} would be of shape $[12,1,8,8]$.
Note that the second element in the shape can be eradicated by a simple reshape,
which does not alter the data representation in memory or its computational
efficiency.

Applying the same reasoning to the $FC$ layer, we conclude that we can
use \code{mconv} again.  Without additional reshapes, the shape of the
layer $S_2$ would be $[12,1,4,4]$.  A fully connected layer is a convolution
with the weight that is identical to the shape of the input array.
Therefore, as we intend to compute ten weighted sums of all the elements,
$W$ now has shape of $[10,12,1,4,4]$.  This yields \code{mconv} to
return a result of shape $[10,1,1,1,1]$.
With these observations it becomes clear that the only parts of
\autoref{cnn} left to complete the implementation are
the pooling layers.

\paragraph{Average Pooling}
The pooling layer $S_1$ can be constructed in a two step process
similarly to the convolution layers. An average pooling of a single image
can be implemented as:
\begin{lstlisting}[language=sac]
float[.,.] avgpool(float[.,.] in) {
   return { iv -> average({ ov -> in[iv*2+ov]
                          | ov < [2,2] })
          | iv < shape(in) / 2 };
}
\end{lstlisting}
We select sub-arrays of shape $[2,2]$ and compute their individual average,
resulting in a matrix of half as many rows and columns as the input.
Based on this definition, a generic version that applies \code{avgpool}
to the two innermost axes of an $n$-dimensional array
can be expressed as:
\begin{lstlisting}[language=sac]
float[*] avgpool(float[*] in) {
   return { iv -> avgpool(in[iv])
          | iv < drop([-2], shape(in)) };
}
\end{lstlisting}
Note that for convenience we overload the name \code{avgpool}.
With these few functions, we are now ready to define the whole network from \autoref{cnn} as
\begin{lstlisting}[language=sac]
float[10,1,1,1,1] forward(
                  float[28,28] I,
                  float[6,5,5] k1, float[6]  b1,
               float[12,6,5,5] k2, float[12] b2,
            float[10,12,1,4,4] fc, float[10] b) {
  c1 = sigmoid(mconv(I, k1, b1));
  s1 = avgpool(c1);
  c2 = sigmoid(mconv(s1, k2, b2));
  s2 = avgpool(c2);
  return sigmoid(mconv(s2, fc, b));
}
\end{lstlisting}
Explicit shapes in \code{forward} are given for documentation
purposes, and they can be replaced with more generic shapes in case of
more abstract networks.

The implementation so far suffices for using the network in \emph{forward mode},
\ie{} once suitable weights and biases are known, we can classify images.
To adjust the weights, we use training inputs where we know the correct answer
for every input image.  The error in recognition is our cost function that we
minimise by using stochastic gradient descent to adjust the weights.

\paragraph{Backpropagating Convolution}
Our loss function has a form $\frac{1}{2}\sum {(y - f(x, w))}^2$, so its derivatives for $w_i$
will have a form $\frac{\partial f}{\partial w_i}f(x, w)\sum y - f(x,w)$
according to the chain rule.   That is, to adjust the weights, we
multiply the error with the derivative of the network with respect to the weights.
The linear nature of the convolution implies that the derivatives are constants, namely the
input of the convolution itself. Consequently, we can approximate the error in the weights
as a convolution of the input with the error:
\begin{lstlisting}[language=sac]
float[*] backweights(float[*] d_out, float[*] in) {
  return conv(in, d_out);
}
\end{lstlisting}
The resulting deltas then can be used to adjust the corresponding weights for the next forward run.
To cater for the imprecision, this is done by applying a factor, usually referred-to as
\textsl{rate}.
\begin{lstlisting}[language=sac]
...
weights = weights - rate*backweights(d_out,in);
...
\end{lstlisting}
Similarly, we can approximate the error of the bias as a sum of the error since the derivative
of the bias is constant 1:
\begin{lstlisting}[language=sac]
float[*] backbias(float[*] d_out) {
  return sum(d_out);
}
\end{lstlisting}
The trickiest bit of implementation is the propagation of the error back to the inputs
of the convolution, which we need to feed into the computation of the next
backpropagation layer.  Mathematically, the derivatives are simply the weights.
The challenge arises from the fact that the outer elements of the result are influenced
by fewer weights than the inner elements.
This distinction between inner elements and outer elements can either be expressed by
embedding the values of the array of errors into a larger
array of zeros or it requires the summations for the boundary elements to range over
fewer products. Here, we opt for the later approach:
\begin{lstlisting}[language=sac]
float[*] backin(float[*] d_out,
                float[*] k, float[*] in) {
  return {
    iv -> off = where(iv<shape(d_out),
                      0*shape(k),
                      iv-shape(d_out)+1);
          sum({ ov -> k[ov+off]
                      * d_out[iv-(ov+off)]
              | ov < min(min(shape(k), iv+1),
                         shape(k)-off) })
    | iv < shape(in)
  };
}
\end{lstlisting}
While all inner elements of the result are computed by \lstinline[breaklines=true,breakatwhitespace = true]{k[ov] * d_out[iv-iv]}
with \code{ov} ranging over the entire shape of the weights
\code{k}, the boundary elements are only computed using those factors where \code{iv-ov}
lies within the shape of \code{d\_out}.
For the boundary elements on the lower end it suffices to restrict the number of products
that are being summed up. This is achieved by the expression
\lstinline{min(shape(k), iv+1)}.
For the elements on the higher end, we only include the products with the higher-indexed
weights. This is achieved by computing an offset vector \code{off}.
For each index \code{iv} beyond the shape of the error in \code{d\_out}, the offset
is set accordingly. This is implemented by using the
standard function \lstinline{where(m, b, c)} which for each \code{iv} chooses
\sel{b}{iv} if \sel{m}{iv} is true and \sel{c}{iv} otherwise\footnote{
    The local binding to \code{off} is not the official
    \sac{} syntax, but we use it for the sake of readability.
    A semantically equivalent formulation of \code{backin} can be found
    at \url{https://github.com/SacBase/CNN}.}.
Finally, we use this offset vector in order to restrict the number of products
for the upper boundaries as well. This is achieved by the outer minimum against
\lstinline{shape(k)-off}.

\paragraph{Backpropagating Average Pooling}
Average Pooling usually is back-propagated by evenly spreading out the error across the indices
that we have averaged across in the forward mode.
In our example, we can express this as:
\begin{lstlisting}[language=sac]
float[.,.] backavgpool(float[.,.] d_out) {
  return { iv -> d_out[iv/2] / 4f
         | iv < shape(d_out) * 2 };
}

float[*] backavgpool(float[*] d_out) {
  return { iv -> backavgpool(d_out[iv])
         | iv < drop([-2], shape(d_out)) };
}
\end{lstlisting}
With these main building blocks, the back-propagation can be implemented in a way very similar
to that of the \code{forward} function shown above. Details can be found at
\url{https://github.com/SacBase/CNN}.

\section{\label{sec:performance}Evaluation}

\pgfplotstableread{%
framework sac tfpy tfpymkl tfcxx tfcxxmkl ptmkl
1 144 102.32 57.87 93.86 50.74 65.89
10 18.2 44.19 27.43 36.80 22.41 29.52
20 12.4 38.57 27.76 30.14 23.03 31.01
30 10.7 38.41 29.45 29.55 22.23 30.71
40 9.4 37.51 30.61 29.42 23.95 38.92
50 7.8 37.73 34.73 30.04 24.09 40.82
60 7.9 38.19 37.54 30.25 25.71 39.78
}{\runtimesta} 

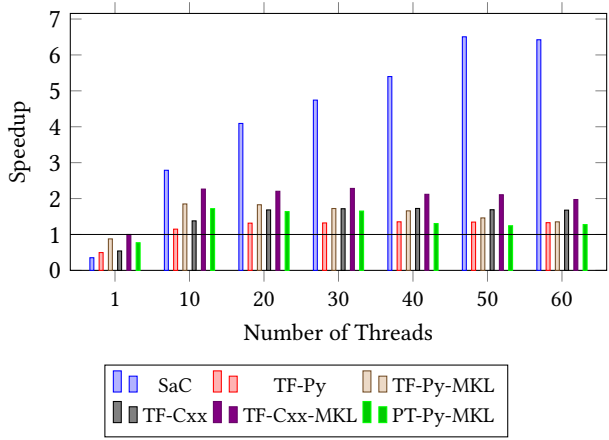
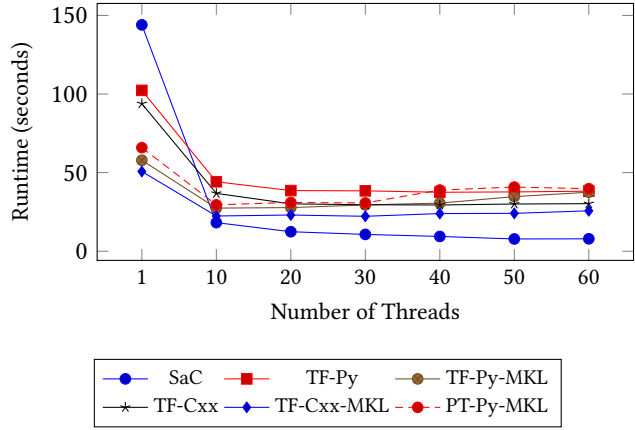
\begin{figure*}[h!]
\begin{subfigure}[t]{0.49\textwidth}
\begin{tikzpicture}
    \begin{axis}[%
        label style={font=\small},
        width=\linewidth,
        height=5cm,
        ybar,
        ylabel={Speedup},
        ymin=0,
        xlabel={Number of Threads},
        xticklabels from table={\runtimesta}{framework},
        xtick=data,xticklabel style={%
            font=\small,
        },
        ytick distance=1,
        bar width={1.5pt},
        legend style={%
            anchor=south,
            at={(0.44,-0.65)},
            legend columns=3,
            font=\footnotesize
        },
        after end axis/.append code={%
            \draw[ultra thin,black] (axis cs:\pgfkeysvalueof{/pgfplots/xmin},1) --
            (axis cs:\pgfkeysvalueof{/pgfplots/xmax},1);
        }]
        \addplot+ table [x expr=\coordindex,y expr={50.74/\thisrow{sac}}] {\runtimesta};
        \addlegendentry{SaC}
        \addplot+ table [x expr=\coordindex,y expr={50.74/\thisrow{tfpy}}] {\runtimesta};
        \addlegendentry{TF-Py}
        \addplot+ table [x expr=\coordindex,y expr={50.74/\thisrow{tfpymkl}}] {\runtimesta};
        \addlegendentry{TF-Py-MKL}
        \addplot+ table [x expr=\coordindex,y expr={50.74/\thisrow{tfcxx}}] {\runtimesta};
        \addlegendentry{TF-Cxx}
        \addplot+ table [x expr=\coordindex,y expr={50.74/\thisrow{tfcxxmkl}}] {\runtimesta};
        \addlegendentry{TF-Cxx-MKL}
        \addplot+ table [x expr=\coordindex,y expr={50.74/\thisrow{ptmkl}}] {\runtimesta};
        \addlegendentry{PT-Py-MKL}
    \end{axis}
\end{tikzpicture}
\captionsetup{width=0.8\linewidth}
\caption{Speedups (higher is better) over the fastest sequential
    runtime \textsc{TF-Cxx-MKL}.}\label{fig:resultss}
\end{subfigure}%
\begin{subfigure}[t]{0.49\textwidth}
\begin{tikzpicture}
    \begin{axis}[%
        label style={font=\small},
        width=\linewidth,
        height=5cm,
        ylabel={Runtime (seconds)},
        xlabel={Number of Threads},
        xticklabels from table={\runtimesta}{framework},
        xtick=data,xticklabel style={font=\small,},
        legend style={%
            anchor=south,
            at={(0.43,-0.65)},
            legend columns=3,
            font=\footnotesize
        }]
        \addplot+ table [x expr=\coordindex,y expr={\thisrow{sac}}] {\runtimesta};
        \addlegendentry{SaC}
        \addplot+ table [x expr=\coordindex,y expr={\thisrow{tfpy}}] {\runtimesta};
        \addlegendentry{TF-Py}
        \addplot+ table [x expr=\coordindex,y expr={\thisrow{tfpymkl}}] {\runtimesta};
        \addlegendentry{TF-Py-MKL}
        \addplot+ table [x expr=\coordindex,y expr={\thisrow{tfcxx}}] {\runtimesta};
        \addlegendentry{TF-Cxx}
        \addplot+ table [x expr=\coordindex,y expr={\thisrow{tfcxxmkl}}] {\runtimesta};
        \addlegendentry{TF-Cxx-MKL}
        \addplot+ table [x expr=\coordindex,y expr={\thisrow{ptmkl}}] {\runtimesta};
        \addlegendentry{PT-Py-MKL}
    \end{axis}
\end{tikzpicture}
\caption{Wallclock Runtimes (lower is better).}\label{fig:resultsr}
\end{subfigure}
\caption{\label{fig:results}CPU only results for \sac{} versus \tf{} and
    \pt{} implementations using up to 64
    Opteron cores, training 10 epochs on 10k images and classifying 10k images.}
\end{figure*}

We now present an evaluation of our CNN implementations in \sac{}, comparing it to
semantically identical implementations in \tf{} and \pt{}\footnote{All three versions can be found in the supplementary material.}.
We discuss programming productivity reflecting our implementation experience,
then we present our experimantal setup for runtime evaluation and we present
performance analisys of the implementations.

\subsection{Effect on Programming Productivity}

Programmer productivity is a very personalised topic as the background in tool familiarity
influences the experience.  The tools that we are comparing are of a
different nature: \sac{} is a general-purpose language, whereas \tf{} and \pt{}
are specifically designed for machine learning purposes, both highly performance-tuned and optimised
for algorithms like the CNN\@.  Neither of these tools
executes the specification directly. Instead, the specification is analysed and translated
into code that executes on parallel architectures.

All three specifications of the CNN in \sac{}, \tf{} and \pt{}
are very similar.
In all systems about 150 lines of code are needed to specify the network and to
orchestrate the reading of inputs and data initialisation.
In all three versions, the programmer needs to understand the abstractions used; in \tf{} and \pt{},
the programmer needs to learn semantics of the available components;
in \sac{}, the programmer needs to understand the building blocks that we described in \autoref{sec:background}.
In \sac{} the backward propagation needs to be specified explicitly,
while the frameworks tools support automatic differentiation.

Further experience is based on the fact that we also had to write the building
blocks in \sac{}, if we assume that the machine learning framework is provided
as an external tool then the next paragraph is not relevant.  Otherwise,
we found the conciseness of the building blocks very satisfying.  The key
components of which are described in \autoref{sec:cnn} can be implemented in
about 150 lines of code.
For someone with reasonable familiarity in \sac{}, this can be achieved within a few hours,
depending on the familiarity with the underlying algorithm.
Overall, the time we spent on the \sac{} implementation was considerably smaller
than the time we needed
to understand sufficient details about the \tf{} and \pt{} frameworks.
Of course this experience is hard to generalise,
but we are rather sure that in cases where the hardware architecture and the machine
learning algorithm is fixed,  figuring out the details about the frameworks and
implementing the algorithm from scratch is very likely to take comparable time.

\subsection{Setup}
Our machine is equipped with 4 AMD Opteron 6376
CPUs (for a total of 64 cores) and an NVIDIA K20 GPU (CUDA driver version
410.79).  We use GCC 7.2.0, \texttt{sac2c} 1.3.3, \cuda{} 10.0,
\python{} 3.6.6, \tf{} 1.12.0 and \pt{} 1.2.0 for all applications.

We compile both frameworks from sources to make sure that the architecture specific flags like
\texttt{-march=native}\linebreak{} \texttt{-mtune=native} are passed to the \textsc{C}/\cxx{} compilers so that
we get proper vectorisation and cost models.  Secondly, \tf{} and \pt{} can
make use of the Intel MKL library~\cite{intel-mkl} to accelerate linear algebra
operations on Intel architectures and provide a significant speedup.
It is not
compiled in by default for \tf{}, and though we could make our comparison
without it, it would not be an honest comparison. Therefore we have verified
that MKL is correctly included in both frameworks, which has made a noticeable
runtime difference.
To make a fair comparison, we
use a \tf{}
with and without Intel MKL~\cite{intel-mkl} activated (indicated by the
\texttt{-MKL} postfix), and we implement the CNN
using both the \python{} and \cxx{} interface (indicated by the \texttt{Cxx}
postfix).  The latter is to check whether static
compilation has an impact on performance. This gives us five framework-based implementations,
plus the on in \sac{}. We run these on both the CPU and GPU, and measure the wall-clock
runtime of the entire application.

With non-MKL \tf{} versions we set the
number of threads via the session variables; for the MKL version, as the library
uses OpenMP, setting the \tf{} threads at the sametime can quickly oversubscribe
the system, so per our experiments the best \tf{}-MKL runtime is achieved when
\tf{} threads are set to 1 and the \texttt{OMP\_NUM\_THREADS} is set to the desirable
value. With \sac{} we control the number of threads by setting
the \texttt{-mt} flag of the binary file, that is automatically created by the
compiler when using the multi-threaded backend.

The applications are run using the following parameters: 10 epochs, 100 images
batch size, 10000 training images and labels, and
10000 test images and labels. The back-propagation has a
learning rate factor of 0.05, and we do not use any momentum.

\subsection{Results and Analysis}
\hyperref[fig:results]{Figure~\ref*{fig:resultss}} shows speedup compared to the
fastest sequential runtime and \autoref{fig:resultsr} shows the runtimes in
seconds.
From \autoref{fig:resultsr} we can see that \sac{} outperforms all the other
frameworks by a noticeable factor ($3\times$ over the best parallel runtime,
$6\times$ over the best sequential runtime), even though being noticeably slower on a
single thread.
From \autoref{fig:resultss} we see that none of the frameworks manage to
achieve more than a $2\times$ speedup. We observe that \tf{} applications using MKL are
up to $2\times$ faster. Additionally, using the \cxx{} over the \python{} interface is
faster by a constant factor.

\begin{table*}
\caption{Best wallclock runtimes (seconds) on the given hardware for each framework.}\label{tab:bestresults}
\centering
\small
\begin{tabularx}{\textwidth}{XXXXXXXXX}
    \toprule
    \textit{Framework} & SaC & TF-Py & TF-Py-MKL & TF-Cxx & TF-Cxx-MKL & PT-Py-MKL \\ \midrule
    \textit{Configuration}  & Opteron 50~threads & NVIDIA K20   & NVIDIA K20   & NVIDIA K20   & NVIDIA K20  & NVIDIA K20 \\ \addlinespace
    \textit{Runtime} & 7.8   & 17.06 & 18.26 & 17.68 & 14.1 & 24.32 \\ \bottomrule
\end{tabularx}
\end{table*}
%
%
%

%

The measurements in \autoref{fig:resultss} show that for 10 threads we have
speedups of up to a factor of 2 for the MKL-based applications, with the
\python{} and \cxx{} only applications achieving a speedup factor of about 1.5.
The \sac{} application has a speedup factor of close to 3. As the number of
threads increases, we see only minor improvements in speedup for the \tf{} and
\pt{} applications, with a best speedup factor of 2.3. The \sac{} application on
the other hand continues to scale, reaching a max speedup factor of 6.4 using 50
threads. From \autoref{fig:resultsr} we can see this runtime plateauing more
clearly. With MKL the runtimes are better than without for the \tf{} and \pt{}
applications, by almost a factor of 2. Additionally, using \cxx{} instead of
\python{} provides slightly better runtimes. We additionally re-ran the CNN
using different batch-sizes and observered no significant change in the the previously
observered scaling.

The speedup plateauing that we see for the \tf{} and \pt{} applications relates
to how nodes within the network graph are translated to threads. Some nodes have
dependencies on outputs of others, and so are scheduled differently to nodes that
have no dependencies. This limits the degree to which work can be distributed
across the threads, affecting the max amount of scaling possible. Changes to the
design of the network, or using a completely different neural-network can lead
to different degrees of scaling. As \sac{} only translates \wloop{}s to
threads, and \wloop{}s are guaranteed to be side-effect free, this leads to
better scaling.

With \tf{} we have two levels of parallelism, as the MKL operations spawn their
own threads independently of the \tf{} scheduler. We tried different
combinations of threading configurations looking for the best possible performance.
Using the default configuration, where \tf{} and MKL use all cores, leads
to oversubscription and degraded performance. Using combinations of values that
match the number of logical cores on the system, such as 16
MKL threads and 4 \tf{} threads, did not lead to better performance compared to
just setting \tf{} thread number to 1 and having MKL scale to all 64 logical
cores. In any case we were not able to resolve the scaling plateau by this
means.


\autoref{tab:bestresults} shows the best runtimes per application and the
hardware configuration this was achieved on. The best runtime for \sac{} is 7.8
seconds using 50 threads. All other applications have their best runtime on the
GPU, with the \tf{} \cxx{} MKL implementation having the best runtime at 14.01 seconds.
The \sac{} application did not perform well on the GPU compared to the other
applications, running $9\times$ slower.

A significant reason for this is the scheduling of communication and also kernel
launches, that are not effectively orchestrated together. For \tf{} and \pt{},
multiple CUDA streams are used to interleave communication and kernel launches
such that they achieve a high degree of latency hiding.

\subsection{Source of Performance in \sac{}}\label{sec:perfw}

The \sac{} compiler is pretty sophisticated, it uses several hundred optimisations
that run in a cycle, therefore explaining what exactly the compiler is doing
to make the code run well is challenging.  It would be good though to identify
the key components, in order to potentially apply the demonstrated capabilities
in other contexts, such as more mainstream function languages.

After looking at intermediate states of the code, we identify the following
necessary optimisations: folding~\cite{sac-folding}, fusion~\cite{sac-fusion},
memory reuse~\cite{sac-memoryreuse}, and statically-scheduled multi-threaded
execution~\cite{sac-mt}.

\paragraph{Folding}
The main idea behind folding is the classical list equality $\map\ f \circ \map\ g =
\map\ f \circ g$, which can the eliminate creation of intermediate arrays.
In the set notation this equality looks like:
\begin{lstlisting}
  a = {iv -> g | iv < u};
  b = {iv -> $\lambda$iv.f a[$K$ iv] | iv < u}
\end{lstlisting}
\noindent
\begin{center}
$\Rightarrow$~%
\begin{tabular}{c}
\begin{lstlisting}
b = {iv -> $\lambda$iv.f (g ($K^{-1}$ iv))  | iv < u}
\end{lstlisting}
\end{tabular}
\end{center}
when $K = \text{identity}$, we get exactly the above equality.  However, very often
$K$s have computable inverses in which case the transformation is applicable to
a larger set of examples.  For the CNN case, we can definitely merge together
the addition of biases and the computation of sigmoid functions in layers $C_1$,
$C_2$ and $FC$.

\paragraph{Fusion}
Here we apply a variant of the classical loop optimisation to
array comprehensions.
The optimisation combines the body of two consequent loops with an identical
iteration space.  In the \map{}-based analogy this would be:
\begin{gather*}
    (\map\ f\ a, \map\ g\ a) = \textit{unzip}\ (\map\ f \Delta\, g\ a)\\
    \text{where}\qquad{}(f \Delta\,g)\ x = (f\ x, g\ x)
\end{gather*}
Even though intuitively, this transformation does not make much sense on lists,
it helps to localize the applications of $f$ and $g$ and share commons
subexpressions.
In \sac{} this optimisation
is quite a bit more sophisticated, as explicit indexing makes it possible to
define fusions on individual partitions of the index-spaces.

\paragraph{Memory Reuse}
This memory analysis makes it possible to do array operations
in-place.  For example, when we increment all the elements by a constant:
\[
    b = \tc{iv}{a[iv] + 1}{\shp{a}}
\]
we can avoid allocating new memory, and reuse $a$, if $a$ is not used
further in the program. The analysis becomes challenging when we consider
reusing existing, but no longer referenced, arrays within the current scope.
The analysis needs to handle conditionals within the set expression or the
access patterns of candidate arrays.

\paragraph{Statically-Scheduled Parallelism}
Finally, our array comprehensions are data parallel by design.  Therefore,
it is relatively straight forward to generate the code that partitions the
index space into chunks and runs each chunk in parallel.  Unfortunately, there
is a lot of small details that makes it very hard to implement this efficiently~\cite{sac-dynadap}.
First of all, one needs to choose the operations we want to run in parallel,
and their granularity.  Secondly, choosing a schedule even for a single array
operation is challenging.  Finally, thread synchronisation and memory management
make a significant difference.  By default we use static scheduling, a custom memory
allocator and for each operation we decide to run in parallel we try to choose
the chunking that maximises the work each active thread is doing.


\section{\label{sec:related}Related Work}

\subsection{Array Languages}
Directly or indirectly, APL~\cite{Iverson:1962:APL} has influenced all existing
array languages. At its core, APL provides a set of operators with a number of
rules on how they can be composed. All operators are either unary or binary,
first- or second-order functions, expressed with a single symbol, which
gives a lot of expressiveness.  
For example, all the building blocks of our CNN
can be expressed in 10 lines of code~\cite{Sinkarovs:2019:CNN:3315454.3329960}.
APL is an untyped language, so all errors
will occur at runtime only.  It comes only with an interpreter and
all the operators are implemented as library functions, limiting cross-operator
optimisations.

Other array languages can be roughly
divided into three groups: direct descendants of APL, grandchildren and further
relatives.  Direct descendants are languages like: J~\cite{jlang}, K~\cite{klang} or
Nial~\cite{nial}. 
They treat every object as an array (maybe except functions) and
provide a large subset of APL operators.  Typically, these languages come
only with interpreters, which limits the optimisations space and performance.
The grandchildren like \sac{}, Futhark~\cite{henriksen2017futhark},
Remora~\cite{remora}, Qube~\cite{TROJAHNER2009643}
are still array-oriented languages, but instead of providing built-in APL
operators natively, they offer a few low-level constructs from which the
operators could be implemented as library functions.  All the mentioned
languages are functional and come with compilers that are
focused on generating high-performance code.  All these languages have
strong static type systems.
Futhark and \sac{} are capable
of generating GPU code automatically.
Finally, further relatives like Matlab~\cite{MathWorks:refguide},
Julia~\cite{julialang},
Python~\cite{python} with Numpy~\cite{numpy} have some notion of
multi-dimensional arrays and a subset of APL operators, both of which
are embedded in the context of the general purpose language.  All the mentioned 
languages come with interpreters only and rarely provide exceptional levels of
performance, yet they are very useful for prototyping.

\subsection{Machine learning DSLs}
Machine Learning DSLs provide a way to express neural-networks using
high-level specifications. Typically, the high-level specification is
either handled by a machine learning framework,
or transformed into machine code for performance reasons.

TypedFlow\footnote{\url{https://github.com/GU-CLASP/TypedFlow}} is embedded
in Haskell and provides a number of dependently-typed primitives that can
be used to define a network.  Later this specification is translated into
\tf{} calls.  This approach provides type safety, powerful
syntax, but performance-wise, it sill relies on the underlying framework.
The tensorflow-ocaml\footnote{\url{https://github.com/LaurentMazare/tensorflow-ocaml}}
and ocaml-torch\footnote{\url{https://github.com/LaurentMazare/ocaml-torch}} are
similar wrappers for \tf{} and \pt{} in Ocaml.

DEFIne~\cite{define-lang} mainly focuses on liberating
data scientists from the necessity to deal with general-purpose languages, such as Python,
when describing the networks.  The proposed syntax focuses exclusively on the
machine learning primitives, and the accompanying tools take care of performance
and portability, still using state of the art machine learning frameworks as a
backend.

DeepDSL~\cite{deep-dsl}, OptiML~\cite{optiml} and Latte~\cite{latte}
focus on optimisations that are specific to machine learning such as
kernel-fusion and parallelisation. They generate code
to C++ and CUDA, using highly-optimised libraries.

The XLA~\cite{xla} is a domain-specific compiler that focuses on accelerating
linear algebra operations in machine-learning applications.  The compiler is a
part of the \tf{} framework, and it works by analysing dataflow graph of
the network and turning it into fast machine code by fusing pipelined nodes, inferring
tensor shapes and performing memory optimisations based on these data sizes.

Tensor Comprehensions~\cite{tensor-comp} has a very similar idea: it is a DSL that is
integrated into existing machine learning frameworks and it provides a common
ground to implement machine learning operators for further cross-optimisation.
A distinctive feature for this approach is the use of the polyhedral model to
perform the actual fusion, blocking, non-trivial scheduling and parallelisation.
In a way the approach is very similar to Halide~\cite{Halide}, except the domain is
different and the number of optimisations is larger.

Diesel~\cite{diesel} is a standalone DSL from NVIDIA that also relies on polyhedral
framework to perform cross-operator optimisations and generate code for CUDA\@.

\subsection{High-Performance Libraries}
Most of the machine learning frameworks rely on highly-optimised libraries that
implement tensor or linear algebra operations.  The
Eigen~\cite{eigen} and Aten\footnote{Available as a part of \pt{} at
    \url{https://github.com/pytorch/pytorch/tree/master/aten}} provide basic tensor operation and are being
used by \tf{} and \pt{} correspondingly.  The MKL~\cite{intel-mkl} and
OpenBlas~\cite{openblas} implement high-performance Basic Linear Aalebra Subrtoutines
(BLAS) for CPUs.  ATLAS~\cite{Whaley:1998:ATL:509058.509096},
BTO~\cite{Belter:2009:AGC:1654059.1654119} and SPIRAL~\cite{doi:10.1177/1094342004041291}
use automatic
tuning to obtain the most efficient implementation of commonly used numerical
algorithms on a chosen architecture.  BLAS operations on CUDA are provided by
CUBLAS~\cite{cublas}.\ cuDNN~\cite{cudnn} implements basic deep learning operations on
GPUs.  NNPACK\footnote{Available at \url{https://github.com/Maratyszcza/NNPACK/}}
and PCL-DNN~\cite{pcldnn} implement deep learning primitives
on CPUs.

\section{\label{sec:conclusions}Conclusions}

This paper makes an argument for an alternative design of machine learning
frameworks.  Instead of using a large number of interconnected specialised
libraries, we consider using one compilable array-oriented language to host
both the framework and the specification of the actual networks.  To justify
the viability of the proposed approach, we implement a minimalistic framework
in native \sac{} and use it to define a state of the art
CNN.  We compare its performance and expressiveness against
\tf{} and \pt{}.

Our solution is concise: about a 150 lines of code to define the
building blocks of the network, and another 150 lines to define the network itself
--- which is about the same amount as for the \tf{} and \pt{} versions.
The basic building blocks in \sac{} are rank-polymorphic functions that
can be easily reused in other contexts.  Rank polymorphism is a key to
expressiveness here. 
By the nature of layered neural networks, the notion of proximity arises
naturally in multiple dimensions: neighbourhood of pixels, neighbouring neurons, connectivity 
between layers, convolutional weights, etc.
In our simple example with 3 layers of neurons ($C_1$, $C_2$ and $FC$ from
Fig.~\ref{cnn}), we ended up with a single definition for multi-dimensional convolution
which could serve all three layers with their varying ranks (3,4, and 5).
The rank-polymorphic nature allows for arbitrarily ranked tensors and thus
can be used for arbitrarily nested networks.
The same holds for the other operations such as the backward propagation.

Our performance experiments on a 64-core machine with a GPU show that the
\sac{} implementation outperforms
\tf{} and \pt{} by a factor of 2 and 3 respectively, even though the chosen
architecture should be ideally suitable for both frameworks.
In particular the performance results came as a big surprise to us, given the
stark difference in implementation efforts. 
As discussed in Section~\ref{sec:performance}, it seems that the interplay
of different tools in \tf{} and \pt{} are getting into the way of achieving
excellent parallel performance, whereas the lean design in the array language setup enables 
better optimisations and ultimately better wallclock runtimes.

When shifting from the domain-specific frameworks to array-oriented languages
we loose the domain-knowledge for optimisation on
the one hand while we gain a unified high-level representation on the other.
At least for the example of CNNs, it seems that the former does not provide
any advantages for the frameworks whereas the latter clearly benefits the array language 
approach.  Also, a unifying representation makes it very easy to extend a framework.
There is no need to understand complexities of the udnerlying design ---
a new building block in the form of a user defined function will be
immediately picked up by a compiler and included in the global program optimisations.



As for productivity and expressiveness, there is no doubt that right now
\python{} is more advanced than any of the existing array languages, at least
in the number of libraries provided by the community.  At the same time,
there seem to be no \emph{conceptual} problem in bringing the same experience and
functionality to the array language of choice.  In the case of \sac{} two
immediate problems will have to be solved: interactive behaviour and
automatic differentiation.

Right now \sac{} is a compiled language and by default we do not get the
same interactivity
as with \python{} in the context of existing machine learning frameworks.
However, right now there exists a jupyter-based frontend that mimics
\python{}-like interactivity.  Several other array languages are more
interactive than \sac{}, and creating an interpreter for \sac{} is
straight-forward.  One then can envision using such an
interpreter for quick prototyping and a compiled version of the same code
for deployment.

Right now \sac{} does not support automatic differentiation which is a
very useful feature that can be found in most of the machine learning
frameworks.  Adding automatic differentiation to a compiler
is a well-understood problem as for example demonstrated by
Stalin\hspace{-1px}$\nabla$~\cite{stalingrad}, therefore bringing it
to the context of an array language is a matter of implementation effort.


By no means do we suggest that existing frameworks can be readily replaced
by array languages.  However, a clean design that eliminates a number of
abstraction layers, supported by the fact that a prototypical research compiler
can significantly outperform two industrial frameworks suggests that the
proposed approach is interesting enough to be further investigated.

\bibliography{paper}

\end{document}